%% file: acl_latex.tex
\documentclass[11pt]{article}

\usepackage[final]{acl}

\usepackage{times}
\usepackage{latexsym}
\usepackage[T1]{fontenc}
\usepackage[utf8]{inputenc}
\usepackage{microtype}
\usepackage{inconsolata}
\usepackage{graphicx}
\usepackage{tikz}
\usepackage{multirow}
\usepackage{multicol}
\usepackage{booktabs}
\usepackage{amsmath}
\usepackage[table]{xcolor}
\usepackage[most]{tcolorbox}
\tcbset{
  aibox/.style={
    width=474.18663pt,
    top=10pt,
    colback=white,
    colframe=black,
    colbacktitle=black,
    enhanced,
    center,
    attach boxed title to top left={yshift=-0.1in,xshift=0.15in},
    boxed title style={boxrule=0pt,colframe=white,},
  }
}
\newtcolorbox{AIbox}[2][]{aibox,title=#2,#1}
\newif\ifarxiv
\arxivtrue

\definecolor{darkred}{rgb}{0.5, 0.0, 0.0}
\definecolor{skyblue}{RGB}{203, 221, 245}

\title{\texttt{LiCoMemory}: Lightweight and Cognitive Agentic Memory for Efficient Long-Term Reasoning}

\author{
 \textbf{Zhengjun Huang\textsuperscript{1}},
 \textbf{Zhoujin Tian\textsuperscript{1}},
 \textbf{Qintian Guo\textsuperscript{1}},
\\
 \textbf{Fangyuan Zhang\textsuperscript{2}},
 \textbf{Yingli Zhou\textsuperscript{3}},
 \textbf{Di Jiang\textsuperscript{4}},
 \textbf{Zeying Xie\textsuperscript{4}},
 \textbf{Xiaofang Zhou\textsuperscript{1}}
\\
 \textsuperscript{1}The Hong Kong University of Science and Technology;
\\
 \textsuperscript{2}Huawei Hong Kong Research Center, Hong Kong;
\\
 \textsuperscript{3}The Chinese University of Hong Kong, Shenzhen;
 \textsuperscript{4}WeBank Co., Ltd, Shenzhen
\\
 \small{
   \href{mailto:zhuangff@connect.ust.hk}{zhuangff@cse.ust.hk},
    ztianaf@cse.ust.hk,
    qtguo@ust.hk,
    zhang.fangyuan@huawei.com
}
\\
\small{
    yinglizhou@link.cuhk.edu.cn,
    dijiang@connect.ust.hk,
    zenxie@webank.com,
    zxf@cse.ust.hk
 }
}

\begin{document}
\maketitle
\begin{abstract}
Large Language Model (LLM) agents exhibit remarkable conversational and reasoning capabilities but remain constrained by limited context windows and the lack of persistent memory. 
Recent efforts address these limitations via external memory architectures, often employing graph-based representations, yet most adopt flat, entangled structures that intertwine semantics with topology, leading to redundant representations, unstructured retrieval, and degraded efficiency and accuracy.
To resolve these issues, we propose \texttt{LiCoMemory}, an end-to-end agentic memory framework for real-time updating and retrieval, which introduces \emph{CogniGraph}, a lightweight hierarchical graph that utilizes entities and relations as semantic indexing layers, and employs temporal and hierarchy-aware search with integrated reranking for adaptive and coherent knowledge retrieval.
Experiments on long-term dialogue benchmarks, LoCoMo and LongMemEval, show that \texttt{LiCoMemory} not only outperforms established baselines in temporal reasoning, multi-session consistency, and retrieval efficiency, but also notably reduces update latency. Our official code and data are available at \url{https://github.com/EverM0re/LiCoMemory}.
\end{abstract}

\input{1-intro}

\input{2-related}

\input{3-method}

\input{4-experiment}

\input{5-conclusion}
\input{6-limitation}

\bibliography{custom}

\input{7-appendix}

\end{document}

%% file: 1-intro.tex
\section{Introduction}

\begin{figure}[t]
  \centering
  \includegraphics[width=\linewidth]{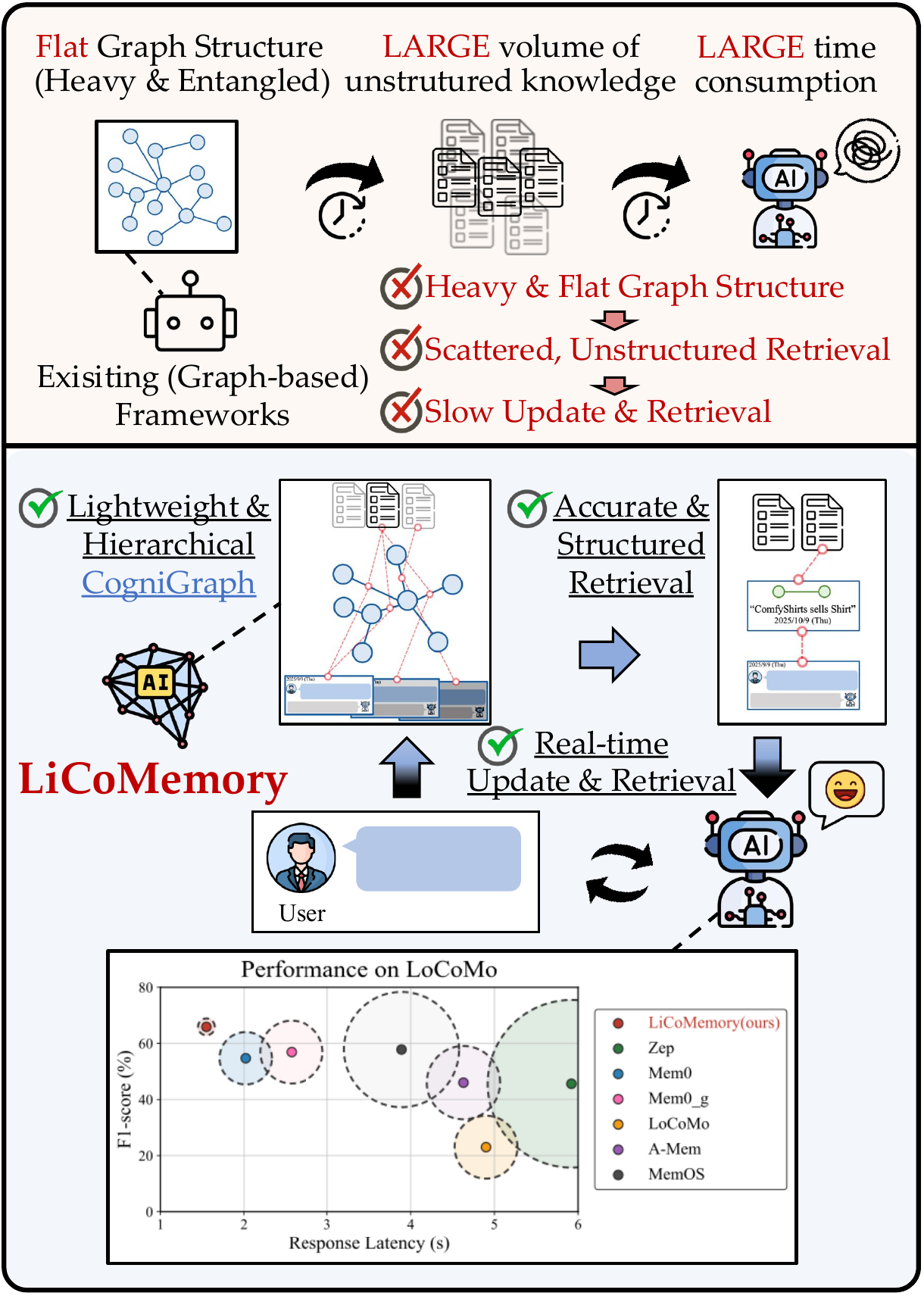}
  \caption{Motivation of \texttt{LiCoMemory}, illustrating how \texttt{LiCoMemory} resolves key challenges of existing memory frameworks. For the performance graph on the bottom, radius of the circles represent the construction token consumption per dialogue.}
  \label{fig:teaser}
\end{figure}

Large Language Models (LLMs) have demonstrated remarkable advancements across a wide range of language understanding and generation tasks~\cite{gpt,qwen} and are increasingly evolving into personalized assistants with enhanced contextual reasoning capabilities~\cite{llmagents}. Despite their strong generalizing and reasoning abilities,  LLMs remain constrained by a critical short-term memory limitation: the finite context window. Information beyond the context window cannot be effectively preserved or recalled, leading to degraded reasoning capability and reduced response accuracy in long-term conversational scenarios~\cite{needmemory}.

To resolve these issues, early attempts to enhance the long-term memory of agents commonly adopt Retrieval-Augmented Generation (RAG) architectures, which leverage conversation history as an external knowledge source and retrieve contextually relevant information to support response generation~\cite{ragagent,ragsurvey}. 
Graph-based RAG further extends this paradigm by structuring conversational content into relational graphs using heuristic rules, thereby capturing semantic dependencies among historical records and improving cross-session reasoning~\cite{graph}. 
Despite their effectiveness in retrieval, these approaches often treat memory as a static component, neglecting the inherently dynamic nature of human–agent interactions, and thus lacking mechanisms for reorganization of accumulated knowledge. 
Moreover, such predefined retrieval and linking strategies can lead to information loss and hinder adaptive memory evolution~\cite{ragagentflaw}.
Recent studies have thus shifted toward dynamic agent memory frameworks, which extend beyond traditional RAG by modeling the evolving nature of conversational data and adaptive retrieval. Mem0~\cite{mem0} initiates this shift by introducing explicit memory operations that allow agents to manage and revise stored knowledge as conversations evolve. 
Building on this, MIRIX~\cite{mirix} enhances retrieval organization through multi-granularity memory indexing and relevance fusion, improving contextual alignment and reducing redundancy in long-term reasoning. Further, Zep~\cite{zep} represents memory as a graph to capture relational dependencies among dialogue events, promoting interpretability but at the cost of high graph-construction overhead and retrieval latency. 

While these frameworks have made progress toward structured conversational memory, some key challenges still remain.
\textbf{(1) Coupled and redundant graph structures:} Existing graph-based memory systems often intertwine semantic content with relational topology, leading to heavy, redundant, and inflexible graph representations that are difficult to adapt to dynamic human–agent interactions~\cite{prob1}. 
\textbf{(2) Scattered and unstructured retrieval:} Owing to flat architectures and unguided retrieval mechanisms, conventional memory pipelines frequently return fragmented or contextually inconsistent information, which undermines reasoning coherence and produces semantically diluted responses~\cite{prob2}. 
\textbf{(3) Slow update and inference:} Large, monolithic graph structures incur substantial computational overhead during incremental updates and inference, limiting their applicability in real-time interaction settings. For instance, GraphRAG~\cite{graphrag} requires up to 20 minutes for graph and community construction per dialogue and over 2 minutes of query latency~\cite{workshop1}.

To address these challenges, we propose \texttt{LiCoMemory}, an end-to-end agentic memory framework that enables real-time updating and retrieval. 
At its core, \texttt{LiCoMemory} introduces \emph{CogniGraph}, a lightweight and semantically aware hierarchical graph that redefines the knowledge graph as a semantic indexing layer rather than a static repository. 
By using graph topology as a structural scaffold instead of embedding extensive content within nodes and edges, CogniGraph indexes and organizes knowledge while linking relational structures to their original textual sources for precise and context-aware reasoning.
During inference, \texttt{LiCoMemory} integrates a unified reranking mechanism that jointly considers semantic similarity, hierarchical structure, and temporal relevance to achieve accurate and structured retrieval. Experimental results demonstrate that \texttt{LiCoMemory} achieves up to 23\% improvement in accuracy over the second-best baseline on established long-term dialogue benchmarks (LoCoMo~\cite{locomo} and LongMemEval~\cite{longmem}), particularly on multi-session and temporal reasoning subsets, while significantly reducing input tokens and response latency, underscoring its efficiency.

Our main contributions are summarized as follows: \textbf{(1) CogniGraph for semantic organization.} We introduce a novel hierarchical graph structure that decouples knowledge storage from semantic organization, transforming the graph into a lightweight, update-friendly semantic index. \textbf{(2) Hierarchy and temporally sensitive retrieval.} \texttt{LiCoMemory }performs structured, top-down retrieval guided by hierarchical relations and temporal cues, ensuring coherent and contextually relevant knowledge selection. \textbf{(3) Efficient and real-time memory operations.} Our lightweight design enables incremental graph construction, fast updates, and low-latency inference during ongoing user--assistant interactions. Together, these components establish \texttt{LiCoMemory} as a unified, real-time memory system capable of retrieving higher-quality, more relevant knowledge and generating contextually grounded responses.

%% file: 2-related.tex
\section{Related Work}

\subsection{Retrieval-Augmented Generation (RAG)}

RAG has emerged as a foundational framework for augmenting LLMs with external memory~\cite{rag1}. A typical RAG pipeline first splits prior interactions or knowledge as segments, then retrieves relevant segments during inference, and provides both the user query and retrieved content to the language model for grounded generation~\cite{depth}.
Due to the highly entangled and semantically redundant nature of conversational data, conventional RAG often retrieves overlapping or loosely related content, resulting in fragmented context and limited reasoning continuity~\cite{ragprob}. To address these limitations, recent research has introduced Graph-based RAG to agent frameworks, where the historical interactions are pre-organized as a graph structure to improve retrieval and reasoning efficiency~\cite{gr1}. 
Beyond flat graph representations, several frameworks further organize memory knowledge hierarchically, such as tree-based structures~\cite{erarag} and community-aware architectures~\cite{archrag}. 
These structured organizations capture cross-session dependencies, reduce redundancy, and enhance contextual coherence, paving the way for more structured and adaptive memory systems.

\subsection{Agent Memory Augmentation}

While LLM-based agents demonstrate remarkable generative capabilities, they remain constrained by limited context windows and the absence of persistent memory, often resulting in inconsistent behavior across extended interactions. 
Agent memory augmentation therefore emerges as a promising direction to address this limitation, aiming to equip conversational agents with external memory systems that support long-term information retention, retrieval, and reasoning. 
LoCoMo~\cite{locomo} introduce a RAG-style conversational framework capable of maintaining multi-session dialogue continuity through chunk-based retrieval and coherence-aware evaluation. 
A-MEM~\cite{amem} advances the concept of agentic memory beyond passive long-term storage by introducing a self-organizing, dynamically evolving memory architecture that autonomously constructs, links, and refines knowledge representations.
LongMemEval~\cite{longmem} further advances this line of research by proposing a dedicated long-term memory model and benchmark for evaluating temporal reasoning, knowledge updating, and cross-session consistency. 
Mem0~\cite{mem0} employs a scalable two-phase architecture (extraction and update) that dynamically stores and retrieves salient facts using a vector database, while also offering a graph-based variant that better supports long term memory maintenance and retrieval.
More recently, Zep~\cite{zep} structures agent memory into knowledge graphs, improving retrieval relevance but suffering from high construction overhead and retrieval latency.

%% file: 3-method.tex
\begin{figure*}[t]
  \centering
  \includegraphics[width=\linewidth]{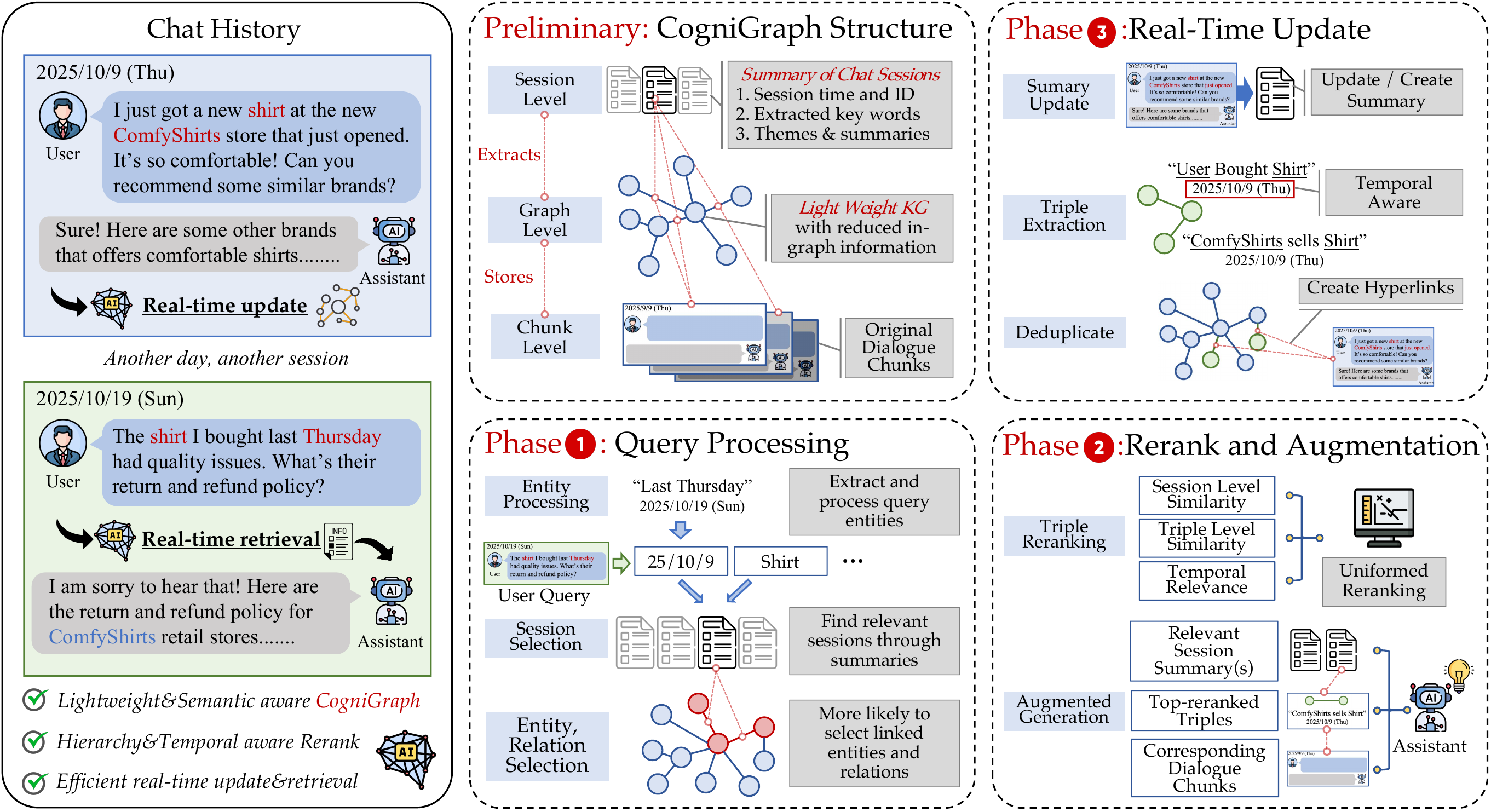}
  \caption{\textbf{Overview of LiCoMemory workflow.} Upon interaction, dialogue chunks are incrementally organized through the CogniGraph (\emph{Preliminary}), a lightweight hierarchical graph linking session summaries, entity–relation triples, and dialogue chunks via cross-layer hyperlinks. New knowledge is continuously integrated and deduplicated to preserve structural consistency (\emph{Phase 1}). At inference, entity extraction and hierarchical retrieval guide top-down search across graph layers (\emph{Phase 2}), followed by hierarchy-temporal–semantic reranking to generate a structured prompt for retrieval-augmented generation (\emph{Phase 3}).
}
  \label{fig:overview}
\end{figure*}

\section{Methodology}
The overall workflow of \texttt{LiCoMemory} is shown in Figure~\ref{fig:overview}. \texttt{LiCoMemory} initiates real-time updates and retrievals during user–assistant interactions. After each dialogue segment, the dialogue chunk with its timestamp and session ID is sent to \texttt{LiCoMemory} for processing, where knowledge is organized and continuously maintained through a lightweight CogniGraph optimized for incremental updates. The system either updates an existing session summary or creates a new one, extracts and deduplicates triples, and establishes cross-level links among session summaries, triples, and dialogue chunks via unique identifiers. At inference time, high-value entities are extracted from user queries in a temporally aware manner to guide top-down retrieval—from session summaries to triples and then to original chunks. Retrieved triples are re-ranked by unified session, triple, and time level relevance, and the resulting summaries, triples, and chunks are integrated into a standard prompt for augmented generation. Further workflow details are provided in the following subsections.


\subsection{CogniGraph: A Lightweight and Semantically-Aware Graph Structure}

Traditional graph-based memory representations often embed extensive semantic content directly within nodes and edges, resulting in entangled representations where structural topology and information content are inseparable. Such designs produce heavy and redundant graphs that hinder efficient updating, leading to unstructured retrieval outputs. 
To address this, we introduce \emph{CogniGraph}, a lightweight and semantically aware hierarchical graph structure that redefines the role of a knowledge graph from a knowledge repository to a semantic indexing layer. Rather than functioning as a storage container for knowledge, CogniGraph employs its graph topology as a structural scaffold that organizes and indexes information across multiple granularities, thereby facilitating efficient retrieval and reasoning.

CogniGraph is composed of three interconnected layers that progressively refine the granularity of information (See Preliminay in Figure~\ref{fig:overview}). \textbf{1) Session level:} Each session node stores a textual summary that captures the high-level context of a user--assistant interaction. 
The summary also contains a set of distilled keywords (\emph{keys}) that represent the central entities, topics, or temporal markers of the dialogue session. 
\textbf{2) Entity-relation level:} This layer constitutes a lightweight knowledge graph composed of entities and relations extracted from dialogue content. 
Each entity node and relation edge retains only essential identifiers without verbose descriptions. 
Entity-relation triples are hyperlinked to their corresponding session summaries, establishing connections between fine-grained semantic relations and their contextual origins. 
\textbf{3) Chunk level:} The lowest layer of CogniGraph stores the original dialogue chunks from which the triples were extracted. Each chunk is also hyperlinked to the triples derived from it, ensuring bidirectional traceability between raw text and structured knowledge. 
From top to bottom, the information granularity increases while structural abstraction decreases, forming a coherent hierarchy that aligns semantics, context, and evidence.

This hierarchical and indexing-oriented design enables CogniGraph to remain compact, easily updatable, and less redundant, while the hyperlink-based cross-layer connections ensure structured and interpretable retrieval. By organizing knowledge as a navigable semantic index rather than an overloaded repository, CogniGraph supports multi-granular reasoning, from abstract contextual understanding to fine-grained evidence retrieval, serving as the structural foundation of \texttt{LiCoMemory}'s retrieval module and enabling efficient, semantically grounded knowledge access for long-term conversational reasoning.

\subsection{Query Processing and Integrated Rerank}
\label{sec:retrieval}
To enable accurate and context-aware retrieval, \texttt{LiCoMemory} adopts a query processing pipeline that combines hierarchical analysis with an integrated temporally aware re-ranking mechanism (Phase 1 and 2 of Figure~\ref{fig:overview}). A user query is first analyzed through entity extraction to identify salient concepts that represent the key information needs of the user. 
The extracted entities are then matched against the summary level of the CogniGraph. 
By comparing the overlap and semantic similarity between the query entities and session summary keys, \texttt{LiCoMemory} ranks all session summaries based on their likelihood of containing relevant information. This process yields a prioritized set of session summaries that serve as entry points for deeper retrieval within the knowledge graph. \texttt{LiCoMemory} then queries the entity-relation level of CogniGraph using the extracted entities as anchors to locate triples that may contain relevant contextual information. Each retrieved triple is associated with its originating session and timestamp, providing both semantic and temporal context. To compute the overall relevance of a triple, \texttt{LiCoMemory} integrates three complementary factors to maintain hierarchy and temporal sensitivity: session-level relevance $S_s$, triple-level relevance $S_t$ and temporal relevance. The unified semantic relevance between the query and a triple is represented by the harmonic mean of the two semantic similarities:
\[
S_{\text{sem}} = \frac{2 S_s S_t}{S_s + S_t},
\]
 To incorporate temporal information without overwhelming semantic relevance, we apply a \emph{Weibull-based decay function} that penalizes outdated triples while retaining a long-tailed contribution for distant timestamps:
\[
w(\Delta\tau) = \exp\!\Big[-\Big(\frac{\Delta\tau}{\hat{\tau}}\Big)^{t_k}\Big],
\quad 0<t_k<1,
\]
where $\Delta\tau$ denotes the time gap between the current query and the triple’s timestamp, and $\hat{\tau}$ is the median of time gaps across retrieved triples, allowing the decay to adapt to the temporal scale of each query. $t_k$ denotes the decay coefficient, where a larger value imposes a stronger penalty on temporally distant information. Finally, the overall relevance score of each triple is defined as
$R(t) = S_{\text{sem}} \times w(\Delta\tau)$, which jointly captures semantic coherence and temporal recency. Because $w(\Delta\tau) \in (0,1]$, the time dimension modulates rather than dominates the semantic signal, ensuring that highly relevant but moderately older information remains retrievable. This formulation achieves a balanced trade-off between temporal adaptivity and semantic precision. The final output is a structured prompt that consolidates highly relevant session summaries, top-ranked triples, and their corresponding original dialogue chunks from which the triples were extracted. This structured representation provides the language model with a coherent and fine-grained view of the relevant knowledge, facilitating precise reasoning across different levels of granularity and enabling robust cross-session understanding.

\subsection{Real-Time Interactions}
\label{met:case}

\begin{figure}[t]
  \centering
  \includegraphics[width=\linewidth]{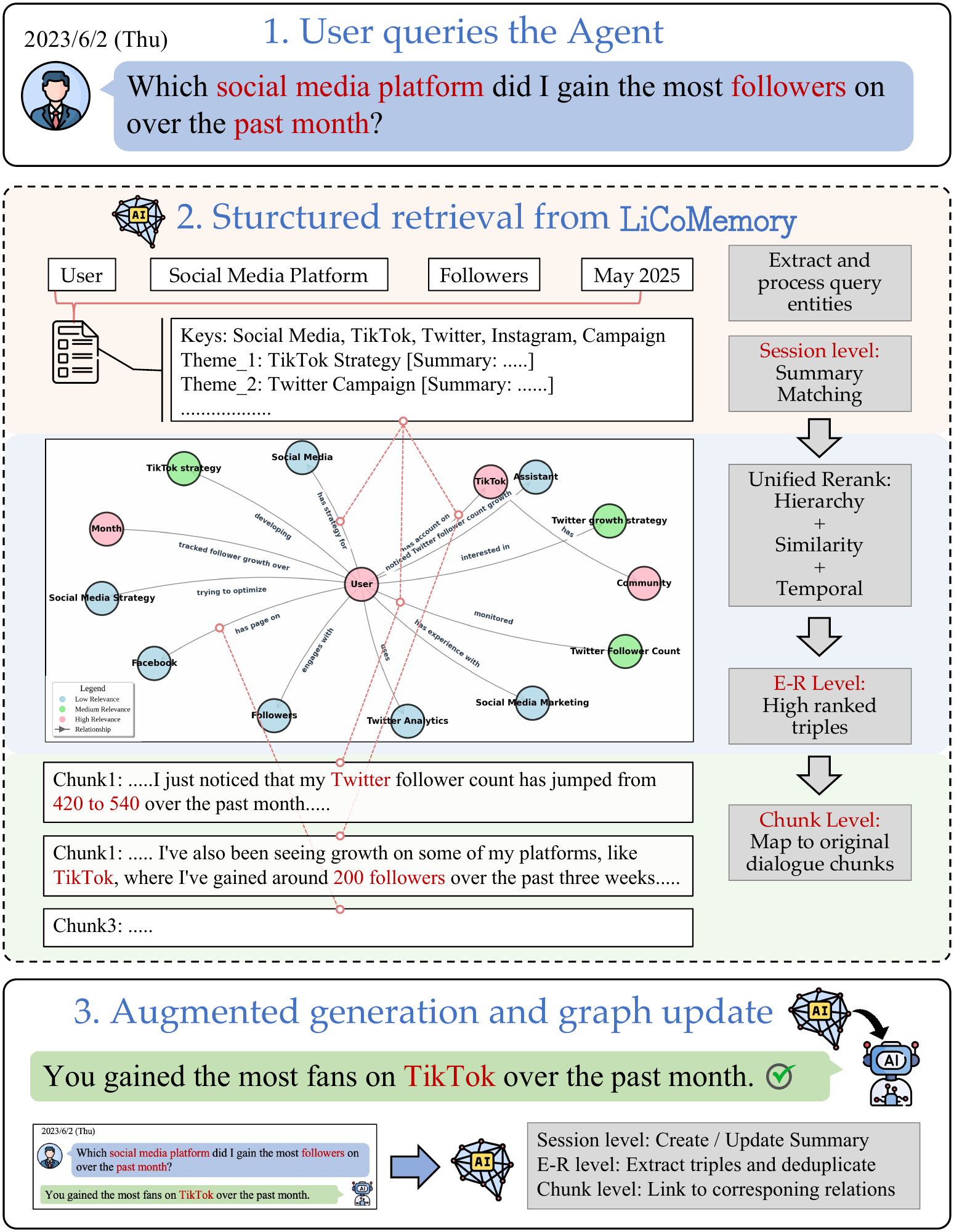}
  \caption{Practical case study of \texttt{LiCoMemory}.}
  \label{fig:case}
\end{figure}

\texttt{LiCoMemory} supports real-time retrievals and updates throughout user-assistant interactions, which consists of two tightly coupled processes. 
First, the agent retrieves relevant knowledge from the existing memory graph and generates a contextually grounded response to the user query. 
Then, \texttt{LiCoMemory} performs real-time memory updates based on both the current interaction and its preceding conversational history, ensuring that the newly acquired information is seamlessly integrated into the hierarchical structure (Phase 3 of Figure~\ref{fig:overview}).

Figure~\ref{fig:case} illustrates a practical example demonstrating how \texttt{LiCoMemory} performs real-time retrieval and memory update during user--assistant interactions. 
As introduced in Section~\ref{sec:retrieval}, the user query is first parsed into structured entities (\emph{e.g.}, Social Media Platform, Followers, Time Period), which guide hierarchical retrieval over the CogniGraph. 
Relevant sessions and triples are then ranked and linked back to their source dialogue chunks, providing grounded evidence for response generation. 
Upon the completion of this interaction, the corresponding dialogue chunk, along with its timestamp and session ID, is transmitted to \texttt{LiCoMemory} for incremental update processing. 
The system first updates the session summary with newly acquired information. 
For ongoing sessions, it refines associated keywords and themes to maintain temporal coherence, 
while for new sessions, it creates a new summary description and its keyword set to represent the contextual core of the dialogue.
Following session-level processing, the system performs triple extraction to transform the current interaction into structured knowledge units.  
The extracted triples are then integrated into the existing entity–relation graph and hyperlinked to their source sessions.
To ensure consistency and eliminate redundancy, we employ type-aware and semantic similarity matching to detect duplicate triples.  
For duplicates, the system links their corresponding sources as additional hyperlinks to the existing nodes instead of creating new ones, thereby maintaining a compact and coherent graph structure. 
Because updates and retrieval share the same CogniGraph backbone, newly added information becomes immediately available for inference without requiring a full re-indexing process. 
Through continuous interaction and memory refinement, the system incrementally maintain a dynamic and temporally consistent representation of user knowledge, ensuring that each response remains contextually coherent and temporally up to date.

%% file: 4-experiment.tex
\section{Experiments}
In this section we evaluate \texttt{LiCoMemory} on real-world datasets to assess its performance. In particular, we aim to answer the following research questions: \textbf{Q1}: How effective is \texttt{LiCoMemory} compared with existing memory paradigms? \textbf{Q2}: How does \texttt{LiCoMemory} perform in real-time practical scenarios? \textbf{Q3}: How does the components of our system affect the final result?

\begin{table*}[t]
  \centering
  \caption{Evaluation on long-term memory QA benchmarks utilizing different language models. The \textbf{best} and \underline{second-best} results are highlighted. $T_R$ stands for query latency while $K_R$ stands for prompt token consumption. All results are averaged over 5 runs and  we report mean $\pm$ standard deviation.}
  \label{tab:qa-performance}
  \renewcommand{\arraystretch}{1}
  \scalebox{0.7}{
  \begin{tabular}{c|c|cccc|ccccc}
    \toprule
    \multirow{2}{*}{\textbf{Model}} & \multirow{2}{*}{\textbf{Method}} 
    & \multicolumn{4}{c|}{\textbf{LongmemEval}} 
    & \multicolumn{4}{c}{\textbf{LoCoMo}} \\
    & & $T_R$ & $K_R$ & Acc. & Rec. & $T_R$ & $K_R$ & Acc. & Rec. \\
    \midrule

    \multirow{8}{*}{\rotatebox[origin=c]{90}{\textbf{\shortstack{Llama-3.1-70B\\Instruct-Turbo}}}} 
    & LoCoMo 
    & 4.51$\pm$1.08s & 3.5$\pm$0.2k & 17.60$\pm$1.05\% & 22.04$\pm$1.12\% 
    & 4.90$\pm$1.42s & 3.2$\pm$0.2k & 23.63$\pm$1.08\% & 25.50$\pm$1.15\% \\
    
    & Memorybank 
    & 8.25$\pm$1.87s & 4.1$\pm$0.2k & 36.40$\pm$0.96\% & 39.21$\pm$1.01\%
    & 7.13$\pm$1.56s & 4.4$\pm$0.2k & 28.80$\pm$0.93\% & 31.52$\pm$0.97\% \\
    
    & MemOS 
    & 3.15$\pm$0.83s & 2.6$\pm$0.2k & 47.80$\pm$0.82\% & 49.03$\pm$0.85\%
    & 3.20$\pm$0.96s & 2.2$\pm$0.2k & 54.10$\pm$0.79\% & 57.53$\pm$0.81\% \\
    
    & Mem0 
    & \underline{1.87$\pm$0.58s} & \underline{2.3$\pm$0.1k} & 56.80$\pm$0.71\% & 61.21$\pm$0.68\%
    & \textbf{1.55$\pm$0.63s} & \underline{2.1$\pm$0.1k} & 53.22$\pm$0.74\% & 57.05$\pm$0.70\% \\
    
    & Mem0$_{g}$ 
    & 2.51$\pm$0.92s & 2.8$\pm$0.2k & 55.40$\pm$0.76\% & \underline{63.09$\pm$0.65\%}
    & \underline{2.11$\pm$0.81s} & 2.4$\pm$0.2k & \underline{55.48$\pm$0.72\%} & \underline{59.32$\pm$0.69\%} \\
    
    & A-Mem 
    & 4.31$\pm$1.26s & 4.5$\pm$0.2k & 57.40$\pm$0.88\% & 62.18$\pm$0.84\%
    & 4.10$\pm$1.68s & 4.2$\pm$0.2k & 43.84$\pm$0.91\% & 49.17$\pm$0.89\% \\
    
    & Zep 
    & 5.22$\pm$1.73s & 4.1$\pm$0.2k & \underline{60.20$\pm$0.81\%} & 62.74$\pm$0.79\%
    & 5.31$\pm$1.94s & 3.8$\pm$0.2k & 40.30$\pm$0.94\% & 51.05$\pm$0.92\% \\
    
    & \cellcolor{skyblue}\textbf{LiCoMemory} 
    & \cellcolor{skyblue}\textbf{1.62$\pm$0.47s}
    & \cellcolor{skyblue}\textbf{1.6$\pm$0.1k}
    & \cellcolor{skyblue}\textbf{69.20$\pm$0.62\%}
    & \cellcolor{skyblue}\textbf{72.39$\pm$0.58\%}
    & \cellcolor{skyblue}\textbf{1.55$\pm$0.59s}
    & \cellcolor{skyblue}\textbf{1.3$\pm$0.1k}
    & \cellcolor{skyblue}\textbf{62.99$\pm$0.71\%}
    & \cellcolor{skyblue}\textbf{64.51$\pm$0.69\%} \\  
    \midrule

    \multirow{8}{*}{\rotatebox[origin=c]{90}{\textbf{GPT-4o-mini}}} 
    & LoCoMo 
    & 5.34$\pm$1.20s & 3.5$\pm$0.2k & 16.60$\pm$1.02\% & 21.56$\pm$1.09\%
    & 4.72$\pm$1.10s & 3.3$\pm$0.2k & 23.87$\pm$1.06\% & 24.91$\pm$1.10\% \\
    
    & Memorybank 
    & 7.93$\pm$1.80s & 4.1$\pm$0.2k & 35.40$\pm$0.95\% & 38.06$\pm$0.98\%
    & 7.62$\pm$1.70s & 4.5$\pm$0.2k & 31.50$\pm$0.92\% & 33.19$\pm$0.96\% \\
    
    & MemOS 
    & 3.72$\pm$0.95s & 2.5$\pm$0.2k & 51.20$\pm$0.78\% & 52.07$\pm$0.81\%
    & 3.96$\pm$1.00s & \underline{2.2$\pm$0.1k} & \underline{58.30$\pm$0.75\%} & 62.93$\pm$0.77\% \\
    
    & Mem0 
    & \underline{1.89$\pm$0.65s} & \underline{2.3$\pm$0.1k} & 62.60$\pm$0.70\% & \underline{71.32$\pm$0.66\%}
    & \underline{1.75$\pm$0.60s} & 2.3$\pm$0.2k & 54.68$\pm$0.73\% & 62.31$\pm$0.71\% \\
    
    & Mem0$_{g}$ 
    & 2.41$\pm$0.85s & 2.9$\pm$0.2k & \underline{64.80$\pm$0.69\%} & 69.53$\pm$0.67\%
    & 2.34$\pm$0.80s & 2.5$\pm$0.2k & 56.96$\pm$0.71\% & \underline{63.14$\pm$0.68\%} \\
    
    & A-Mem 
    & 4.52$\pm$1.40s & 4.3$\pm$0.2k & 55.00$\pm$0.86\% & 59.30$\pm$0.84\%
    & 4.63$\pm$1.50s & 4.1$\pm$0.2k & 48.59$\pm$0.88\% & 53.82$\pm$0.86\% \\
    
    & Zep 
    & 6.12$\pm$1.60s & 4.2$\pm$0.2k & 58.60$\pm$0.83\% & 61.02$\pm$0.80\%
    & 5.92$\pm$1.50s & 3.7$\pm$0.2k & 44.76$\pm$0.91\% & 46.51$\pm$0.93\% \\
    
    & \cellcolor{skyblue}\textbf{LiCoMemory} 
    & \cellcolor{skyblue}\textbf{1.74$\pm$0.55s}
    & \cellcolor{skyblue}\textbf{1.7$\pm$0.1k}
    & \cellcolor{skyblue}\textbf{73.80$\pm$0.60\%}
    & \cellcolor{skyblue}\textbf{76.63$\pm$0.57\%}
    & \cellcolor{skyblue}\textbf{1.61$\pm$0.50s}
    & \cellcolor{skyblue}\textbf{1.2$\pm$0.1k}
    & \cellcolor{skyblue}\textbf{67.20$\pm$0.69\%}
    & \cellcolor{skyblue}\textbf{68.09$\pm$0.67\%} \\
    
    \bottomrule
  \end{tabular}
  }
\end{table*}

\subsection{Experimental Setup}

\noindent\begin{tikzpicture}
\filldraw (0,0) -- (-0.15,0.08) -- (-0.15,-0.08) -- cycle; 
\end{tikzpicture} \textbf{Dataset.} The performance of \texttt{LiCoMemory} was evaluated on two public long-term memory benchmarks: \textbf{LongMemEval} and \textbf{LoCoMo}. 
\textbf{LongMemEval}~\cite{longmem} is a comprehensive benchmark for evaluating long-term memory in conversational agents, consists of 500 questions across six types: single-session user (S.S.U.), single-session assistant (S.S.A.), single-session preference (S.S.P.), multi-session, temporal reasoning, and knowledge update.
\textbf{LoCoMo}~\cite{locomo} focuses on extremely long multi-session dialogues, containing 1,986 questions in five distinct categories: single-hop, multi-hop, temporal, open-domain and adversial reasoning. 
Detailed statistics can be found in Appendix~\ref{app:data}.

\noindent\begin{tikzpicture}
\filldraw (0,0) -- (-0.15,0.08) -- (-0.15,-0.08) -- cycle; 
\end{tikzpicture} \textbf{Metrics.} We evaluate different methods from two perspectives: response quality and efficiency.
Following prior work~\cite{longmem}, response quality is measured using Accuracy (Acc.) and Recall (Rec.), with the evaluation prompts detailed in Appendix~\ref{app:prompt}.
Accuracy is assessed using the LLM-as-a-Judge protocol from LongMemEval~\cite{longmem}, in which a large language model performs binary judgments of answer correctness. As partial credit is not permitted, this metric provides a more faithful, human-aligned estimate of retrieval quality.
For the Adversarial subset of the LoCoMo dataset, where ground-truth answers are unavailable, all responses are labeled as ``Context insufficient to answer.''
Recall is defined as the proportion of ground-truth targets retrieved within the top-15 items.
Efficiency is evaluated in terms of token consumption and query latency~\cite{mem0}, corresponding to the total number of tokens used by the LLM during query processing and the time required to generate a complete response.

\noindent\begin{tikzpicture}
\filldraw (0,0) -- (-0.15,0.08) -- (-0.15,-0.08) -- cycle; 
\end{tikzpicture} \textbf{Baselines.} 
We compare \texttt{LiCoMemory} with several well-established baselines, including 
LoCoMo~\cite{locomo},
Zep~\cite{zep},
Mem0(Mem0$_g$)~\cite{mem0},
A-Mem~\cite{amem},
Memorybank~\cite{Memorybank},
and MemOS~\cite{memos}.
Detailed introduction is listed in Appendix~\ref{app:baseline}.

\noindent\begin{tikzpicture}
\filldraw (0,0) -- (-0.15,0.08) -- (-0.15,-0.08) -- cycle; 
\end{tikzpicture} \textbf{Implementation Details.} 
All methods are evaluated under the same settings. Llama-3-8B-Instruct serves as the primary LLM for memory construction, while BGE-M3 is adopted for text embedding to support retrieval. 
During answer generation, Llama-3.1-70B-Instruct-Turbo and GPT-4o-mini are utilized as generation models. All experiments are conducted on NVIDIA 80G A100 GPUs. We report results averaged over 5 independent runs to ensure reliable performance and runtime measurements. The number of retrieved memory units ($k$) is set to 15 and the decay coefficient ($t_k$) is set to 0.1 with details shown in Appendix~\ref{app:parameter}.

\subsection{Main Results(RQ1)}
Table~\ref{tab:qa-performance} presents a comprehensive comparison of \texttt{LiCoMemory} with representative memory frameworks across two long-term dialogue benchmarks using different backbone language models. Across both Llama-3.1-70B-Instruct-Turbo and GPT-4o-mini, \texttt{LiCoMemory} consistently achieves the highest accuracy and recall while maintaining the lowest or near-lowest query latency.
Moreover, its performance remains stable across multiple runs, demonstrating strong robustness.
Specifically, on LongMemEval, it surpasses the second-best baseline by 9.0\% in accuracy and 9.3\% in recall with Llama-3.1-70B, and by 9.0\% and 5.3\% respectively with GPT-4o-mini. Similarly, on LoCoMo, LiCoMemory outperforms Mem0$_g$ by 7.5\% in accuracy and 5.2\% in recall on Llama-3.1-70B, and achieves a 8.9\% and 4.95\% gain under GPT-4o-mini. The observed improvements in QA performance demonstrate the effectiveness of the proposed CogniGraph structure. In addition, the reductions in retrieval latency and retrieval volume further substantiate the advantages of its precise retrieval mechanism and lightweight, efficiency-oriented graph design. Notably, the most pronounced performance gains are observed on the LoCoMo benchmark, where retrieval latency is reduced by 10\% and token consumption by 45\% compared to the second-best baseline (Mem0).
 
\begin{figure}[t]
  \centering
  \includegraphics[width=\linewidth]{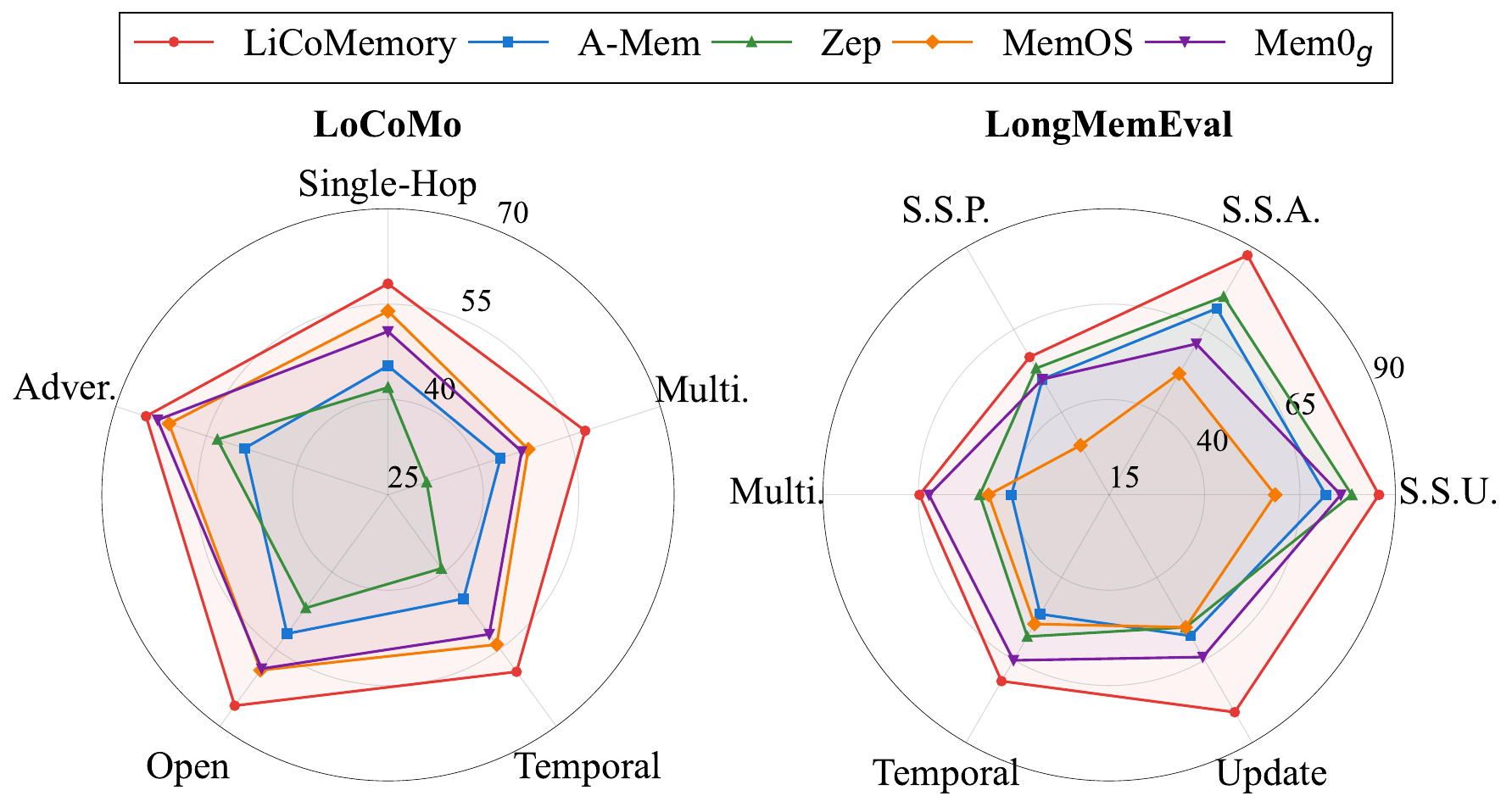}
  \caption{Accuracy breakdown of \texttt{LiCoMemory} and baselines on subsets of LoCoMo and LongmemEval. }
  \label{fig:radar}
\end{figure}

To further analyze how \texttt{LiCoMemory} outperforms other frameworks on long-term dialogue benchmarks, we provide a detailed breakdown of its performance across subsets with different focuses using GPT-4o-mini as the generation model, as shown in Figure~\ref{fig:radar}. As illustrated in the left chart, \texttt{LiCoMemory} consistently surpasses the second-best baseline (MemOS) across all subsets of LoCoMo, with a large gain observed in the Temporal-Reasoning subset, where accuracy improves by 19.2\%.
A smaller improvement is observed on the Adversarial subset, likely due to occasional false positives arising when the correct entries are not retrieved, as reflected by the recall results in Table~\ref{tab:qa-performance}. On the LongMemEval benchmark, \texttt{LiCoMemory} achieves substantial gains on the Multi-Session (26.6\%) and Temporal Reasoning (15.9\%) subsets compared to the second best baseline (Mem0), highlighting the effectiveness of our CogniGraph structure and unified reranking mechanism in capturing temporal and cross-session dependencies. 

\begin{figure*}[t]
  \centering
  \includegraphics[width=\linewidth]{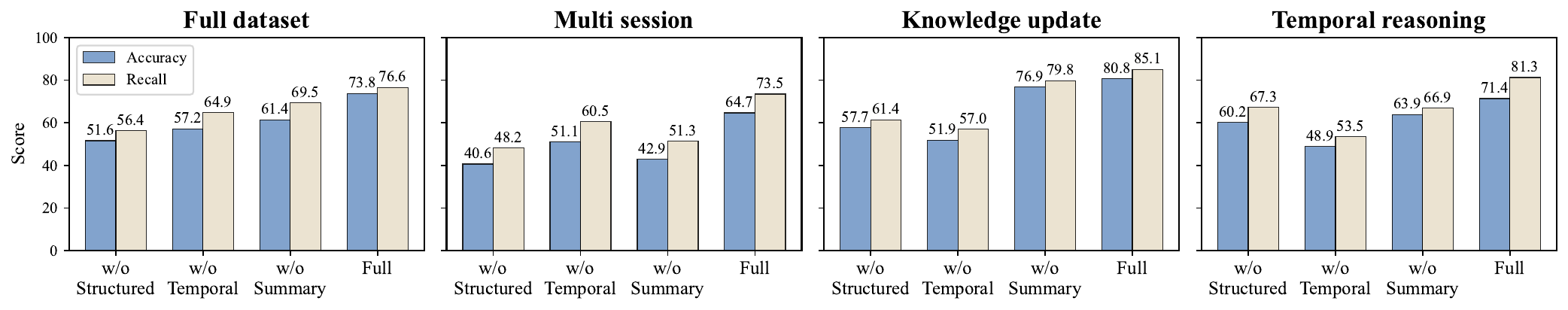}
  \caption{Ablation study of \texttt{LiCoMemory} on LongmemEval and subset breakdown. }
  \label{fig:ablation}
\end{figure*}

\subsection{Real-Time Performance(RQ2)}

Following the discussion in Section~\ref{met:case}, we evaluate \texttt{LiCoMemory} in a practical interactive setting where the agent must support real-time updates and retrieval. Using Llama-3.1-70B-Instruct-Turbo as the backbone, we perform chunk-by-chunk insertion on the LoCoMo dataset to emulate real-world conversational flows. During context ingestion, we only insert dialogue chunks without triggering retrieval, and issue queries after all insertions are completed. As shown in Table~\ref{tab:real}, \texttt{LiCoMemory} maintains leading accuracy with minimal degradation from static to real-time insertion, while achieving the lowest token usage and latency in both context processing ($K_G$, $T_G$) and querying ($K_R$, $T_R$). With the support of CogniGraph, \texttt{LiCoMemory} further reduces construction token cost by over 3 times and construction latency by more than an order of magnitude, without compromising retrieval quality. These results highlight both the efficiency benefits introduced by CogniGraph and the robustness of \texttt{LiCoMemory} for real-time interactive deployment.

\begin{table}[t]
  \centering
  \caption{Detailed performance of \texttt{LiCoMemory} and baselines on LoCoMo in real-time interaction. $K_G$ stands for token consumption per session during context processing stage while $K_R$ stands for prompt token input. Accordingly, $T_G$ stands for latency per session of context processing stage $T_R$ stands for query latency.}
  \label{tab:real}
  \renewcommand{\arraystretch}{1.1}
  \scalebox{0.75}{
  \begin{tabular}{c|c|cc|cc}
    \toprule
    \multirow{2}{*}{\textbf{Method}} & \multirow{2}{*}{\textbf{Accuracy}} 
    & \multicolumn{2}{c|}{\textbf{Token}} 
    & \multicolumn{2}{c}{\textbf{Latency}} \\
    & & $K_G$ & $K_R$ & $T_G$ & $T_R$ \\
    \midrule
    \textbf{Zep} & 38.7\% & 212.5k & 4.0k & 2871s & 5.71s\\
    \textbf{Mem0} & 54.68\% & 49.3k & 2.2k & 1772s & 1.78s\\
    \textbf{Mem0$_{g}$} & 55.82\% & 61.8k & 2.4k & 2081s & 2.25s\\
    \textbf{A-Mem} & 44.12\% & 30.7k & 4.1k & 209s & 4.63s\\
    \textbf{MemOS} & 54.08\% & 143.2k & 2.6k & 256s & 2.71s\\
    \midrule
    \cellcolor{skyblue}\textbf{LiCoMemory} & \cellcolor{skyblue}66.4\% & \cellcolor{skyblue}13.52k & \cellcolor{skyblue}1.3k & \cellcolor{skyblue}21s & \cellcolor{skyblue}1.52s\\
    
    \bottomrule
  \end{tabular}
  }
\end{table}

\subsection{Ablation Study(RQ3)}

We analyze the contribution of each major component of \texttt{LiCoMemory} through an ablation study, where individual modules are selectively removed to assess their impact across diverse evaluation scenarios. Figure~\ref{fig:ablation} summarizes the results, showing that disabling different components leads to varying degrees of performance degradation, thereby revealing their complementary roles in supporting coherent long-term reasoning, temporal consistency, and effective cross-session retrieval.

Overall, all ablated variants exhibit clear performance drops, with the most severe degradation observed for \textbf{w/o Structured retrieval}. When retrieval is performed solely over extracted triples, ignoring session hierarchy and entity--relation structure, performance drops from 73.8/76.6 to 51.6/56.4 on the full dataset and from 64.7/73.5 to 40.6/48.2 in the multi-session setting, indicating fragmented retrieval and weakened factual grounding. Removing temporal weighting (\textbf{w/o Temporal awareness}) causes sharp declines in time-sensitive tasks, with temporal reasoning falling from 71.4/81.3 to 48.9/53.5 and knowledge update from 80.8/85.1 to 51.9/57.0, demonstrating the necessity of temporal signals for avoiding outdated evidence. Finally, disabling summary-level guidance (\textbf{w/o Summary}) consistently degrades performance (73.8/80.6 $\rightarrow$ 61.4/69.5 on the full dataset; 64.7/73.5 $\rightarrow$ 42.9/51.3 in multi-session settings), as retrieval becomes overly local and fails to capture higher-level contextual coherence. Together, these results confirm that structured retrieval, temporal awareness, and summary-level abstraction jointly underpin the robustness of \texttt{LiCoMemory}.

%% file: 5-conclusion.tex
\section{Conclusion}
This paper presents \texttt{LiCoMemory}, an end-to-end agentic memory framework designed for real-time updating, retrieval, and reasoning in long-term conversational scenarios. \texttt{LiCoMemory} incorporates CogniGraph, a lightweight and semantically aware hierarchical graph structure that redefines the role of knowledge graphs as a semantic indexing layer rather than a static repository. By leveraging hierarchical and temporally sensitive retrieval, the system unifies session-level, relational-level, and temporal relevance to retrieve coherent and contextually aligned knowledge. Experimental results on long-term dialogue benchmarks demonstrate that \texttt{LiCoMemory} consistently retrieves quality information and achieves superior performance in temporal and multi-session reasoning and other complicated tasks, while significantly improving update efficiency and inference speed compared to existing baselines. In future work, we plan to extend our structure to multi-agent settings and explore adaptive memory compression strategies to further enhance scalability and reasoning capability.

%% file: 6-limitation.tex
\section{Limitations}

The current LoCoMo framework is limited to single-modality conversational data.
Although it effectively models long-term structure within text-based interactions, it cannot incorporate additional modalities such as images, audio signals or structured sensor data.
This restriction narrows its applicability in real-world settings where multimodal grounding is essential for maintaining coherent memory across heterogeneous inputs.

Another limitation lies in the LLM-dependent graph construction process.
Building and refining the memory graph requires invoking large language models for abstraction, relation inference, and coherence evaluation.
This dependence on large language models is not unique to LoCoMo but represents a broader challenge shared by many LLM-based memory organization methods.
As the volume of conversational history scales up, the number of required model calls grows correspondingly, leading to substantial computational and monetary cost.
This poses practical challenges for deploying LoCoMo in large-scale or high-throughput applications.

%% file: 7-appendix.tex
\appendix

\section{Experiment Details}

\subsection{Dataset Details}
\label{app:data}

We conduct experiments on two publicly available long-term dialogue datasets, \textbf{LongMemEval}~\cite{locomo} and \textbf{LoCoMo}~\cite{locomo}, both designed to evaluate memory-intensive conversational reasoning.

LongMemEval dataset contains 500 dialogues paired with 500 evaluation questions.
Each dialogue comprises an average of 50.22 sessions and 9.83 turns per session, resulting in approximately 101k tokens per dialogue.
The questions are annotated into six categories—single-session-user, multi-session, single-session-preference, temporal-reasoning, knowledge-update, and single-session-assistant—with average lengths ranging from 10 to 36 tokens.

LoCoMo dataset includes 1,986 questions constructed from 10 long-form dialogues.
The dialogues contain an average of 27.2 sessions, each with about 21 turns, yielding roughly 16k tokens per dialogue.
Questions are grouped into five reasoning types, corresponding to Single Hop, Multi Hop, Open Domain, Temporal and Adversarial questions, with average question lengths between 10 and 13 tokens across categories.

\begin{table*}[t]
\centering
\small
\renewcommand{\arraystretch}{1.25}
\setlength{\tabcolsep}{10pt}

\begin{tabular}{l c c}
\toprule
\textbf{Statistic} & \textbf{LongMemEval} & \textbf{LoCoMo} \\
\midrule
Number of Questions & 500 & 1,986 \\
Number of Dialogues & 500 & 10 \\
Avg. Sessions / Dialogue & 50.22 & 27.20 \\
Avg. Turns / Session & 9.83 & 21.63 \\
Avg. Tokens / Dialogue & 101{,}781 & 15{,}965 \\

\midrule
\textbf{Question Type Distribution} &
\begin{tabular}{@{}l@{}}
single-session-user: 70 \\
multi-session: 133 \\
single-session-preference: 30 \\
temporal-reasoning: 133 \\
knowledge-update: 78 \\
single-session-assistant: 56
\end{tabular}
&
\begin{tabular}{@{}l@{}}
Type 1: 282 \\
Type 2: 321 \\
Type 3: 96 \\
Type 4: 841 \\
Type 5: 446
\end{tabular}
\\

\midrule
\textbf{Avg. Tokens per Question (by type)} &
\begin{tabular}{@{}l@{}}
single-session-user: 10.64 \\
multi-session: 14.95 \\
single-session-preference: 16.40 \\
temporal-reasoning: 18.84 \\
knowledge-update: 14.18 \\
single-session-assistant: 36.79
\end{tabular}
&
\begin{tabular}{@{}l@{}}
Type 1: 9.81 \\
Type 2: 10.82 \\
Type 3: 12.28 \\
Type 4: 12.76 \\
Type 5: 12.68
\end{tabular}
\\

\bottomrule
\end{tabular}

\caption{Dataset statistics of LongMemEval and LoCoMo.}
\label{tab:dataset}
\end{table*}

\subsection{Baseline Details}
\label{app:baseline}

In this section, we provide extended descriptions of the baseline systems compared against \texttt{LiCoMemory}. 

\noindent \textbf{LoCoMo}~\cite{locomo} is a long-context reasoning and memory evaluation framework designed to test an agent’s capacity to perform retrieval over extended conversational histories. The system treats the entire multi-session dialogue as an unstructured text corpus, which is then segmented into fixed-length or semantically coherent textual chunks. At inference time, queries are embedded and matched against these chunks using conventional vector similarity search. The top-k retrieved chunks are fed into an LLM to generate a final answer. This pipeline reflects a \emph{classical RAG-style architecture} and does not maintain explicit entity-level structures or temporal links. Instead, LoCoMo emphasizes broad coverage of historical information and robustness across diverse conversational phenomena such as preference tracking, multi-hop reasoning across sessions, and temporal change. Due to its reliance on chunk-based retrieval, the framework is sensitive to chunk granularity and dense-retrieval bottlenecks, particularly when dialogues exceed hundreds of thousands of tokens.

\noindent \textbf{Zep}~\cite{zep} is a retrieval-based conversational memory system that introduces a more structured and selective memory management paradigm. Unlike purely vector-based chunk retrieval, Zep employs \emph{schema-guided memory} with typed memory entries representing specific categories such as user facts, preferences, tasks, temporal events, and environment states. Each memory entry includes metadata such as timestamps, semantic tags, and importance scores, enabling prioritization and temporal filtering. Zep's memory controller supports operations such as \textit{add}, \textit{update}, \textit{expire}, and \textit{recall}, allowing it to perform complex reasoning over temporally extended dialogues. This structured approach enables stronger performance on long-horizon tasks where simple chunk retrieval is insufficient, though Zep’s schema rigidity can limit adaptability in open-domain interactions.

\noindent \textbf{Mem0}~\cite{mem0} provides a modular memory framework designed for deployment in interactive agents that accumulate personal, episodic, and task-specific knowledge over time. The \emph{non-graph} version (denoted as \textbf{Mem0}) represents memory as a set of independent text entries, each stored as an LLM-generated summary or atomic fact. Memory operations are performed via in-context instructions: the system “reflects” over recent conversation to decide whether an event should be added to memory, updated, or ignored. Retrieval is executed through embedding-based nearest-neighbor search over memory entries, using similarity metrics to select relevant items for grounding the agent’s responses. This version is lightweight, easy to integrate, and scalable for applications such as personal assistants or autonomous agent loops. However, because it lacks explicit structural constraints, the system may experience memory redundancy or drift when large numbers of entries accumulate. The graph-based variant (denoted as \textbf{Mem0$_{g}$}) extends the original design by organizing memory into an evolving knowledge graph. Instead of isolated entries, memories are represented as nodes—users, entities, preferences, events—and edges encode explicit relations such as temporal transitions, dependencies, and causal associations. Memory updates may introduce new nodes, refine attributes of existing nodes, or modify relationships to maintain global consistency. Retrieval is conducted through graph traversal, relation-aware embedding, or hybrid neural-symbolic queries. This structure greatly improves multi-hop reasoning, eliminates redundant memory entries, and provides better long-term coherence. Because the graph enforces explicit relational grounding, the graph-based Mem0 is generally more robust on temporally dependent tasks and cross-session preference tracking, though it incurs higher computational overhead and requires controller logic to maintain graph consistency during updates.

\noindent \textbf{A-MEM}~\cite{amem} introduces a dynamically evolving memory architecture designed to capture both short-term conversational signals and long-term user-specific information. A-MEM maintains a layered memory hierarchy that includes:  
(1) a \textit{local memory} for contextual, short-horizon reasoning;  
(2) a \textit{global memory} for long-term facts and user information; and  
(3) a \textit{cross-event relational layer} that links semantically related memory pieces into a coherent structure.  
The system refines stored knowledge via iterative LLM-based consolidation, reducing noise and improving abstraction. During retrieval, A-MEM leverages both content similarity and structural dependencies to select relevant memory components. Its dynamic update mechanism enables continuous refinement of knowledge representations, making it suitable for evolving multi-session environments. However, the reliance on recurrent consolidation steps can introduce latency and occasional over-abstraction of fine-grained details.

\noindent \textbf{MemoryBank}~\cite{Memorybank} is a long-term memory framework designed to endow conversational agents with persistent, user-centric memory across extended interactions. The system organizes memory as a collection of natural-language records summarizing user profiles, preferences, historical events, and interaction traces. These records are incrementally updated through LLM-driven reflection mechanisms that periodically condense recent conversations into concise memory entries. At inference time, MemoryBank retrieves relevant memories via embedding-based similarity search and injects them into the prompt to ground response generation. By emphasizing user personalization and longitudinal consistency, MemoryBank demonstrates strong performance in scenarios requiring stable preference tracking and long-term user modeling. However, because memory entries are stored primarily as free-form text without explicit relational structure, the framework may suffer from redundancy, semantic overlap, and limited multi-hop reasoning capability as memory size grows.

\noindent \textbf{MemOS}~\cite{memos} proposes a memory-augmented conversational architecture that focuses on selective memory formation and efficient retrieval under long-context settings. The system employs an LLM-based controller to determine which conversational elements are worth storing, transforming them into concise memory units that capture salient facts or events. These units are indexed using dense embeddings and retrieved on demand to support downstream reasoning. Unlike static memory accumulation, MemOS emphasizes controlled memory growth by filtering low-utility or redundant information, thereby mitigating memory bloat and retrieval noise. This design enables efficient scaling to long interaction histories while maintaining response relevance. Nevertheless, similar to other embedding-centric approaches, MemOS lacks explicit entity-level or relational representations, which can limit its effectiveness on tasks requiring structured reasoning over temporally or causally linked memories.

\subsection{Prompt Details}
\label{app:prompt}
The following prompts (shown in Fig.~\ref{fig:prompt}) are employed for the LLM-as-a-Judge evaluation protocol introduced by~\cite{longmem}. Different query types correspond to distinct evaluation prompts tailored to their reasoning requirements. 
For Temporal-Reasoning queries, minor off-by-one errors are disregarded to eliminate ambiguity regarding whether the first day is counted. 
For the Single-Session-Preference subset, responses are judged based on their alignment with user preferences inferred from the dialogue history. 
To ensure consistency with the metric defined in the original LoCoMo research~\cite{locomo}, a unified default evaluation procedure is applied across all LoCoMo subsets for fair comparison.

\begin{figure}[t]
  \centering
  \includegraphics[width=\linewidth]{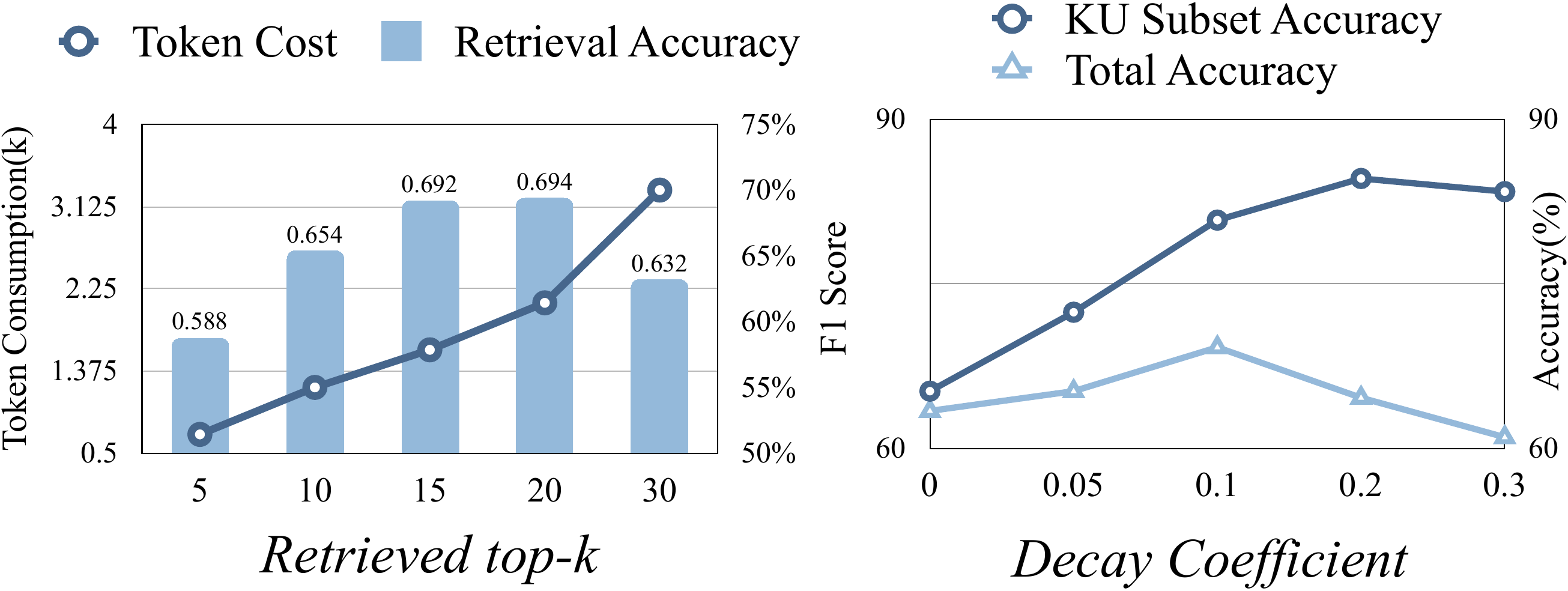}
  \caption{Results of hyperparameter study. From \underline{Left} to \underline{Right}, two graphs in order demonstrates the effect of number of retrieved memory units($k$) and decay coefficient ($t_k$) on \texttt{LiCoMemory}'s performance.}
  \label{fig:hyper}
\end{figure}

\subsection{Hyperparameter Details}
\label{app:parameter}

In this section we evaluate the effect of hyperparameters on the performance of \texttt{LiCoMemory} with respect to the following key parameters: $k$ (number of retrieved memory units), $t_k$ (decay coefficient).

\noindent\begin{tikzpicture}
\filldraw (0,0) -- (-0.15,0.08) -- (-0.15,-0.08) -- cycle ; 
\end{tikzpicture} \textbf{Number of retrieved memory units top-$k$.} The choice of the top-$k$ retrieval parameter is inherently dependent on the characteristics of the target dataset. Selecting a value of $k$ that is too small risks discarding critical information during the retrieval stage, particularly for large-scale datasets, whereas an excessively large $k$ may introduce substantial redundancy and noise, thereby obscuring relevant evidence and hindering the LLM’s ability to identify correct information. Prior memory-centric frameworks evaluated on datasets such as LoCoMo and LongMemEval typically adopt $k$ values in the range of 10 to 30. Motivated by these observations, we conduct an ablation study on LongMemEval to examine the impact of different $k$ values on retrieval accuracy and token consumption during inference. As illustrated in the left panel of Fig.~\ref{fig:hyper}, increasing $k$ leads to a gradual rise in the number of retrieved tokens. Notably, choosing an excessively small value (e.g., $k=5$) results in a substantial degradation in accuracy. This phenomenon can be attributed to the large corpus size of LongMemEval, where insufficient retrieval coverage disproportionately affects multi-session reasoning scenarios. Conversely, performance also declines when $k$ becomes overly large, which may stem from LLM hallucination or the model’s failure to effectively attend to relevant information amid excessive retrieved context. Balancing retrieval effectiveness and computational efficiency, we therefore select $k=15$ as the default setting for all subsequent experiments.

\noindent\begin{tikzpicture}
\filldraw (0,0) -- (-0.15,0.08) -- (-0.15,-0.08) -- cycle ; 
\end{tikzpicture} \textbf{Decay coefficient $t_k$.} Intuitively, a larger $t_k$ imposes a stronger penalty on older information, thereby biasing retrieval toward more recent content. While this mechanism is beneficial for scenarios requiring rapid knowledge updates, it may adversely affect tasks that rely on long-term or cross-session dependencies. To systematically analyze this trade-off, we conduct a hyperparameter study on LongMemEval by varying $t_k$ and evaluating its impact on retrieval accuracy across different subsets. The experimental results shown on the right graph of Fig~\ref{fig:hyper} indicate that, as $t_k$ increases, accuracy on the knowledge update subset improves monotonically and eventually saturates, reflecting the higher relevance of recent information in this setting. However, this property does not generalize to other subsets, where newly introduced information is not necessarily more informative than earlier context. In these cases, an excessively large decay coefficient causes the retrieval mechanism to overemphasize recent but irrelevant memories, leading to a degradation in overall performance. Quantitatively, we observe that increasing $t_k$ within the range of $[0, 0.1]$ consistently improves the overall accuracy, as the benefits on knowledge update scenarios outweigh the mild losses elsewhere. Beyond this range, however, further increasing $t_k$ results in a sharp drop in accuracy, suggesting that overly aggressive temporal decay disrupts effective long-term retrieval. Taking both robustness and task diversity into consideration, we fix the decay coefficient to $t_k = 0.1$ for all subsequent experiments.

\begin{figure*}[h]
\begin{AIbox}{Prompt template for LLM-as-a-Judge.}

{\bf Single-Session-User, Single-Session-Assistant, Multi-Session:} \\
I will give you a question, a correct answer, and a response from a model. Please answer yes if the response contains the correct answer. Otherwise, answer no. If the response is equivalent to the correct answer or contains all the intermediate steps to get the correct answer, you should also answer yes. If the response only contains a subset of the information required by the answer, answer no. \\
Question: \{\textcolor{darkred}{Question}\} 
Correct Answer: \{\textcolor{darkred}{Golden Answer}\}
Model Response: \{\textcolor{darkred}{Model Response}\}\\
Is the model response correct? Answer yes or no only.

{\bf Temporal-Reasoning:} \\
I will give you a question, a correct answer, and a response from a model. Please answer yes if the response contains the correct answer. Otherwise, answer no. If the response is equivalent to the correct answer or contains all the intermediate steps to get the correct answer, you should also answer yes. If the response only contains a subset of the information required by the answer, answer no. In addition, do not penalize off-by-one errors for the number of days. If the question asks for the number of days/weeks/months, etc., and the model makes off-by-one errors (e.g., predicting 19 days when the answer is 18), the model's response is still correct. \\
Question: \{\textcolor{darkred}{Question}\} 
Correct Answer: \{\textcolor{darkred}{Golden Answer}\}
Model Response: \{\textcolor{darkred}{Model Response}\}\\
Is the model response correct? Answer yes or no only.

{\bf Knowledge-Update:} \\
I will give you a question, a correct answer, and a response from a model. Please answer yes if the response contains the correct answer. Otherwise, answer no. If the response contains some previous information along with an updated answer, the response should be considered as correct as long as the updated answer is the required answer.\\
Question: \{\textcolor{darkred}{Question}\} 
Correct Answer: \{\textcolor{darkred}{Golden Answer}\}
Model Response: \{\textcolor{darkred}{Model Response}\}\\
Is the model response correct? Answer yes or no only.

{\bf Single-Session-Preference:} \\
I will give you a question, a rubric for desired personalized response, and a response from a model. Please answer yes if the response satisfies the desired response. Otherwise, answer no. The model does not need to reflect all the points in the rubric. The response is correct as long as it recalls and utilizes the user's personal information correctly.\\
Question: \{\textcolor{darkred}{Question}\} 
Rubic: \{\textcolor{darkred}{Evaluation Rubic}\}
Model Response: \{\textcolor{darkred}{Model Response}\}\\
Is the model response correct? Answer yes or no only.

{\bf LoCoMo:} \\
I will give you a question, a correct answer, and a response from a model. Please answer yes if the response contains the correct answer. Otherwise, answer no. If the response is equivalent to the correct answer or contains all the intermediate steps to get the correct answer, you should also answer yes. If the response only contains a subset of the information required by the answer, answer no. \\
Question: \{\textcolor{darkred}{Question}\} 
Correct Answer: \{\textcolor{darkred}{Golden Answer}\}
Model Response: \{\textcolor{darkred}{Model Response}\}\\
Is the model response correct? Answer yes or no only.

\end{AIbox} 
\caption{Prompt template for LLM-as-a-Judge.}
\label{fig:prompt}
\end{figure*}



%% file: custom.bib
@article{erarag,
  title={EraRAG: Efficient and Incremental Retrieval Augmented Generation for Growing Corpora},
  author={Zhang, Fangyuan and Huang, Zhengjun and Zhou, Yingli and Guo, Qintian and Li, Zhixun and Luo, Wensheng and Jiang, Di and Fang, Yixiang and Zhou, Xiaofang},
  journal={arXiv preprint arXiv:2506.20963},
  year={2025}
}

@inproceedings{MemoryBank,
  author       = {Wanjun Zhong and
                  Lianghong Guo and
                  Qiqi Gao and
                  He Ye and
                  Yanlin Wang},
  title        = {MemoryBank: Enhancing Large Language Models with Long-Term Memory},
  booktitle    = {{AAAI}},
  pages        = {19724--19731},
  publisher    = {{AAAI} Press},
  year         = {2024}
}

@article{memos,
  author       = {Zhiyu Li and
                  Shichao Song and
                  Hanyu Wang and
                  Simin Niu and
                  Ding Chen and
                  Jiawei Yang and
                  Chenyang Xi and
                  Huayi Lai and
                  Jihao Zhao and
                  Yezhaohui Wang and
                  Junpeng Ren and
                  Zehao Lin and
                  Jiahao Huo and
                  Tianyi Chen and
                  Kai Chen and
                  Kehang Li and
                  Zhiqiang Yin and
                  Qingchen Yu and
                  Bo Tang and
                  Hongkang Yang and
                  Zhi{-}Qin John Xu and
                  Feiyu Xiong},
  title        = {MemOS: An Operating System for Memory-Augmented Generation {(MAG)}
                  in Large Language Models},
  journal      = {CoRR},
  volume       = {abs/2505.22101},
  year         = {2025}
}

@article{depth,
  title={In-depth Analysis of Graph-based RAG in a Unified Framework},
  author={Zhou, Yingli and Su, Yaodong and Sun, Youran and Wang, Shu and Wang, Taotao and He, Runyuan and Zhang, Yongwei and Liang, Sicong and Liu, Xilin and Ma, Yuchi and others},
  journal={arXiv preprint arXiv:2503.04338},
  year={2025}
}

@article{archrag,
  title={ArchRAG: Attributed Community-based Hierarchical Retrieval-Augmented Generation},
  author={Wang, Shu and Fang, Yixiang and Zhou, Yingli and Liu, Xilin and Ma, Yuchi},
  journal={arXiv preprint arXiv:2502.09891},
  year={2025}
}

@article{workshop1,
  title={Scalable Graph-based Retrieval-Augmented Generation via Locality-Sensitive Hashing},
  author={Zhang, Fangyuan and Huang, Zhengjun and Zhou, Yingli and Guo, Qintian and Luo, Wensheng and Zhou, Xiaofang},
  journal={Proceedings of the VLDB Endowment. ISSN},
  volume={2150},
  pages={8097}
}

@article{locomo,
  title={Evaluating very long-term conversational memory of llm agents},
  author={Maharana, Adyasha and Lee, Dong-Ho and Tulyakov, Sergey and Bansal, Mohit and Barbieri, Francesco and Fang, Yuwei},
  journal={arXiv preprint arXiv:2402.17753},
  year={2024}
}

@article{longmem,
  title={Longmemeval: Benchmarking chat assistants on long-term interactive memory},
  author={Wu, Di and Wang, Hongwei and Yu, Wenhao and Zhang, Yuwei and Chang, Kai-Wei and Yu, Dong},
  journal={arXiv preprint arXiv:2410.10813},
  year={2024}
}

@article{mem0,
  title={Mem0: Building production-ready ai agents with scalable long-term memory},
  author={Chhikara, Prateek and Khant, Dev and Aryan, Saket and Singh, Taranjeet and Yadav, Deshraj},
  journal={arXiv preprint arXiv:2504.19413},
  year={2025}
}

@article{zep,
  title={Zep: A temporal knowledge graph architecture for agent memory, 2025},
  author={Rasmussen, Preston and Paliychuk, Pavlo and Beauvais, Travis and Ryan, Jack and Chalef, Daniel},
  journal={URL https://arxiv. org/abs/2501.13956}
}

@article{amem,
  title={A-mem: Agentic memory for llm agents},
  author={Xu, Wujiang and Mei, Kai and Gao, Hang and Tan, Juntao and Liang, Zujie and Zhang, Yongfeng},
  journal={arXiv preprint arXiv:2502.12110},
  year={2025}
}

@inproceedings{rag1,
  title={A survey on rag meeting llms: Towards retrieval-augmented large language models},
  author={Fan, Wenqi and Ding, Yujuan and Ning, Liangbo and Wang, Shijie and Li, Hengyun and Yin, Dawei and Chua, Tat-Seng and Li, Qing},
  booktitle={Proceedings of the 30th ACM SIGKDD conference on knowledge discovery and data mining},
  pages={6491--6501},
  year={2024}
}

@article{gr1,
  title={From local to global: A graph rag approach to query-focused summarization},
  author={Edge, Darren and Trinh, Ha and Cheng, Newman and Bradley, Joshua and Chao, Alex and Mody, Apurva and Truitt, Steven and Metropolitansky, Dasha and Ness, Robert Osazuwa and Larson, Jonathan},
  journal={arXiv preprint arXiv:2404.16130},
  year={2024}
}

@inproceedings{needmemory,
  title={Memory matters: The need to improve long-term memory in llm-agents},
  author={Hatalis, Kostas and Christou, Despina and Myers, Joshua and Jones, Steven and Lambert, Keith and Amos-Binks, Adam and Dannenhauer, Zohreh and Dannenhauer, Dustin},
  booktitle={Proceedings of the AAAI Symposium Series},
  volume={2},
  number={1},
  pages={277--280},
  year={2023}
}

@article{llmagents,
  title={Personal llm agents: Insights and survey about the capability, efficiency and security},
  author={Li, Yuanchun and Wen, Hao and Wang, Weijun and Li, Xiangyu and Yuan, Yizhen and Liu, Guohong and Liu, Jiacheng and Xu, Wenxing and Wang, Xiang and Sun, Yi and others},
  journal={arXiv preprint arXiv:2401.05459},
  year={2024}
}

@article{qwen,
  title={Qwen technical report},
  author={Bai, Jinze and Bai, Shuai and Chu, Yunfei and Cui, Zeyu and Dang, Kai and Deng, Xiaodong and Fan, Yang and Ge, Wenbin and Han, Yu and Huang, Fei and others},
  journal={arXiv preprint arXiv:2309.16609},
  year={2023}
}

@article{gpt,
  title={Gpt-4 technical report},
  author={Achiam, Josh and Adler, Steven and Agarwal, Sandhini and Ahmad, Lama and Akkaya, Ilge and Aleman, Florencia Leoni and Almeida, Diogo and Altenschmidt, Janko and Altman, Sam and Anadkat, Shyamal and others},
  journal={arXiv preprint arXiv:2303.08774},
  year={2023}
}

@article{ragagent,
  title={Agentic retrieval-augmented generation: A survey on agentic rag},
  author={Singh, Aditi and Ehtesham, Abul and Kumar, Saket and Khoei, Tala Talaei},
  journal={arXiv preprint arXiv:2501.09136},
  year={2025}
}

@article{graph,
  title={Graph-based Approaches and Functionalities in Retrieval-Augmented Generation: A Comprehensive Survey},
  author={Zhu, Zulun and Huang, Tiancheng and Wang, Kai and Ye, Junda and Chen, Xinghe and Luo, Siqiang},
  journal={arXiv preprint arXiv:2504.10499},
  year={2025}
}

@article{ragagentflaw,
  title={A survey of personalization: From rag to agent},
  author={Li, Xiaopeng and Jia, Pengyue and Xu, Derong and Wen, Yi and Zhang, Yingyi and Zhang, Wenlin and Wang, Wanyu and Wang, Yichao and Du, Zhaocheng and Li, Xiangyang and others},
  journal={arXiv preprint arXiv:2504.10147},
  year={2025}
}

@article{ragsurvey,
  title={A survey on the memory mechanism of large language model-based agents},
  author={Zhang, Zeyu and Dai, Quanyu and Bo, Xiaohe and Ma, Chen and Li, Rui and Chen, Xu and Zhu, Jieming and Dong, Zhenhua and Wen, Ji-Rong},
  journal={ACM Transactions on Information Systems},
  volume={43},
  number={6},
  pages={1--47},
  year={2025},
  publisher={ACM New York, NY}
}

@article{prob1,
  title={You Don't Need Pre-built Graphs for RAG: Retrieval Augmented Generation with Adaptive Reasoning Structures},
  author={Chen, Shengyuan and Zhou, Chuang and Yuan, Zheng and Zhang, Qinggang and Cui, Zeyang and Chen, Hao and Xiao, Yilin and Cao, Jiannong and Huang, Xiao},
  journal={arXiv preprint arXiv:2508.06105},
  year={2025}
}

@article{ragprob,
  title={A survey of personalization: From rag to agent},
  author={Li, Xiaopeng and Jia, Pengyue and Xu, Derong and Wen, Yi and Zhang, Yingyi and Zhang, Wenlin and Wang, Wanyu and Wang, Yichao and Du, Zhaocheng and Li, Xiangyang and others},
  journal={arXiv preprint arXiv:2504.10147},
  year={2025}
}

@article{prob2,
  title={When to use graphs in rag: A comprehensive analysis for graph retrieval-augmented generation},
  author={Xiang, Zhishang and Wu, Chuanjie and Zhang, Qinggang and Chen, Shengyuan and Hong, Zijin and Huang, Xiao and Su, Jinsong},
  journal={arXiv preprint arXiv:2506.05690},
  year={2025}
}

@article{graphrag,
  title={From local to global: A graph rag approach to query-focused summarization},
  author={Edge, Darren and Trinh, Ha and Cheng, Newman and Bradley, Joshua and Chao, Alex and Mody, Apurva and Truitt, Steven and Metropolitansky, Dasha and Ness, Robert Osazuwa and Larson, Jonathan},
  journal={arXiv preprint arXiv:2404.16130},
  year={2024}
}

@article{mirix,
  title={Mirix: Multi-agent memory system for llm-based agents},
  author={Wang, Yu and Chen, Xi},
  journal={arXiv preprint arXiv:2507.07957},
  year={2025}
}
